\documentclass[journal,twoside]{ieeetran}
\usepackage{cite}
\usepackage{amsmath,amssymb,amsfonts}
\usepackage{algorithmic}
\usepackage{graphicx}
\usepackage{textcomp}
\usepackage[caption=false,font=footnotesize]{subfig}

\newcommand{\sign}{\text{sign}}
\newcommand{\thetaapp}{\theta_\text{app}}
\newcommand{\psiapp}{\psi_\text{app}}
\usepackage[hyperref]{xcolor}
\definecolor{ao(english)}{rgb}{0.0, 0.5, 0.0}
\usepackage{hyperref}
\hypersetup{colorlinks, breaklinks, citecolor=blue, linkcolor=ao(english), urlcolor=blue}
\usepackage[nameinlink,noabbrev,sort&compress]{cleveref}
\Crefname{equation}{}{}
\crefname{figure}{Fig.}{Figs.}

\newtheorem{problem}{Problem}
\newtheorem{remark}{Remark}

\newtheorem{definition}{Definition}

\def\BibTeX{{\rm B\kern-.05em{\sc i\kern-.025em b}\kern-.08em
		T\kern-.1667em\lower.7ex\hbox{E}\kern-.125emX}}
\begin{document}
	\title{Docking of Autonomous Vehicles with a Stationary Docking Station in 3D Space}
	\author{Ram Milan Kumar Verma,  Shashi Ranjan Kumar, \IEEEmembership{Senior Member, IEEE}, and Hemendra Arya
		\thanks{R. M. K. Verma, S. R. Kumar, and H. Arya are with the Department of Aerospace Engineering, Indian Institute of Technology Bombay, Powai, Mumbai 400076, India (e-mails: rmverma@aero.iitb.ac.in, srk@aero.iitb.ac.in, arya@aero.iitb.ac.in).}}
	
	\maketitle
	
	\begin{abstract}
		
		In this paper, we present a strategy for autonomous docking of autonomous vehicles in three-dimensional space. Docking is a safety-critical task and requires expert piloting skills. Vehicles with autonomous docking capabilities are highly desirable in various applications, such as marine vehicle docking, aerial vehicle docking, spacecraft docking, and landing. To dock autonomously with the docking station, the vehicle must align itself to a specific desired orientation relative to the docking station and also reduce speed as it approaches. The vehicle achieves near-zero speed to dock successfully and safely without colliding with the docking station. Inspired by the philosophies from the guidance literature, we present a finite-time sliding mode-based strategy to achieve the same. The range and line-of-sight kinematics relations describing the motion of the vehicle with respect to the stationary docking station are used to steer the vehicle to achieve the desired orientation for docking.
		This docking strategy is validated in MATLAB\textsuperscript{\textregistered} simulations for various initial locations and orientations of both the vehicle and the docking station.
	\end{abstract}
	
	
	\section{Introduction}
	\label{sec:introduction}
	\IEEEPARstart{A}{utonomous} vehicles (AV) are being widely adopted for solving various challenging problems pertaining to operations on water, ground, air, and space. 
	These AVs are being extensively utilized across domains such as research, industries, scientific explorations, and defense applications. Many such applications include navigating autonomously to a specified location and docking at the destination.  
	In maritime applications, autonomous underwater vehicles (AUVs) are used for search, inspection of subsea installations, or mapping operations that, upon mission completion, require them to navigate back and dock with the mother ship.
	Also, due to limited energy capacity, the AUVs need to return for recharging, refueling, or tool swaps, etc., without human intervention. 
	This paper addresses the problem of autonomous docking in three-dimensional space.
	
	In general, waypoint- and trajectory-generation-based methods were used to guide the AV to the docking station (DS) at the desired orientation. In \cite{10.1109/TCST.2024.3385666}, the authors used a trajectory tracking control for a fully actuated surface vessel while performing automated docking. The work in \cite{10.1016/j.oceaneng.2018.01.114} implemented a range-only localization algorithm to approach the DS, which decomposed the tasks into two phases, namely: the homing phase and the docking phase. These waypoint- and trajectory-based methods required careful placement of waypoints and generation of trajectories, which can be tedious. 
	The authors in \cite{10.1016/j.oceaneng.2015.08.029} developed a funnel-type docking system for a cabled ocean observatory network. 
	In \cite{10.1016/j.oceaneng.2015.10.015}, the vision positioning was done using two cameras. However, the methods discussed relied on different sensors and control strategies across mission phases. This also required switching between the guidance and control strategies. Also, more sensors increase complexity, cost, and the requirement for onboard computation power.
	
	The concept of docking can also be used for unmanned aerial vehicles (UAVs) \cite{10.1109/LRA.2025.3643302} to land, recharge, or swap batteries, and for station-keeping.
	In \cite{10.1109/TCST.2018.2866963}, the authors used artificial potential-based methods for docking of spacecrafts. \cite{10.1109/LCSYS.2025.3632763,10.1109/TCST.2014.2379639} proposed model predictive control for the rendezvous of spacecrafts. The authors in \cite{10.1109/LCSYS.2021.3136813} used a control barrier function to guarantee safe spacecraft docking.  \cite{10.1016/j.oceaneng.2022.112634} addressed the three-dimensional docking of an AUV. Therefore, autonomous docking is useful for applications across domains, utilizing different vehicles such as spacecrafts, UAVs, mobile robots, and AUVs. The increased number of deployed offshore systems (infrastructures, sensors for sea-health monitoring, water gliders, and others) increases the need for small AVs to detach, perform routine tasks such as periodic inspection and maintenance, and dock autonomously with the DS. The handbook \cite{10.1007/978-3-319-16649-0_16} discussed the practical challenges for autonomous docking of AUVs, such as methods for the AUV to find the dock, to physically attach, to recharge batteries, to establish a communication link, to wait in a low power state for a new mission, and to undock. Many underwater exploration and oceanographic research efforts are constrained by limited battery capacity.  The works in \cite{10.1109/JOE.2016.2554679} discussed USV-based launch and recovery systems for AUVs. One constructive method to solve this problem was to use docking to enable underwater recharging \cite{10.1109/UT.2017.7890282} and data communication.  
	
	Recognizing the importance of docking and its associated challenges, we propose a guidance-based approach that utilizes relative motion kinematics to achieve successful docking. In the interceptor guidance literature, many works, such as \cite{10.1177/0954410016641442,10.1016/j.ast.2021.106735,10.1007/s11071-024-09564-1,10.1016/j.ast.2024.109657}, used a guidance-based strategy to intercept a target at a specific angle. However, they used a constant-speed strategy because the interceptors lack axial thrust control, so there is no active speed control mechanism. The authors in \cite{10.2514/1.G009394} proposed a guidance-based approach to the docking of USVs, which only considers planar motion. In light of the above discussion, the main contributions of this work are summarized below: 
	\begin{itemize}
		\item We propose a nonlinear autonomous docking strategy in three-dimensional space, derived within a nonlinear framework. This allows the proposed strategy to remain valid in a wide envelope. 
		\item Unlike existing works, as in \cite{10.1016/j.oceaneng.2015.10.015,10.1016/j.oceaneng.2015.08.029}, which uses different guidance laws in homing phase and docking phase or uses trajectory generation methods (as in \cite{10.1109/LRA.2025.3643302}), the proposed work presents a unified framework for navigating as well as docking with DS. 
		\item Another advantage of the proposed strategy is that it does not require separate sensors for navigation and docking phases. This also eliminates the need to switch between different control strategies.
		\item Apart from maintaining the specified orientation, the proposed strategy also facilitates a reduction in AV's speed to near zero for a safe docking with the DS. In addition, the proposed strategy relies on range and LOS-related measurements, which are relatively easier to obtain in practice.
	\end{itemize}
	\section{Problem Formulation}
	In what follows, we first define the frames of reference used to describe the equations of motion of AV. Thereafter, the 3-D engagement scenario is presented with respect to the AV's position, the location of DS, and the orientation of the line joining them, also known as the line-of-sight (LOS).
	\begin{figure}[!h]
		\centering
		\includegraphics[width=\linewidth]{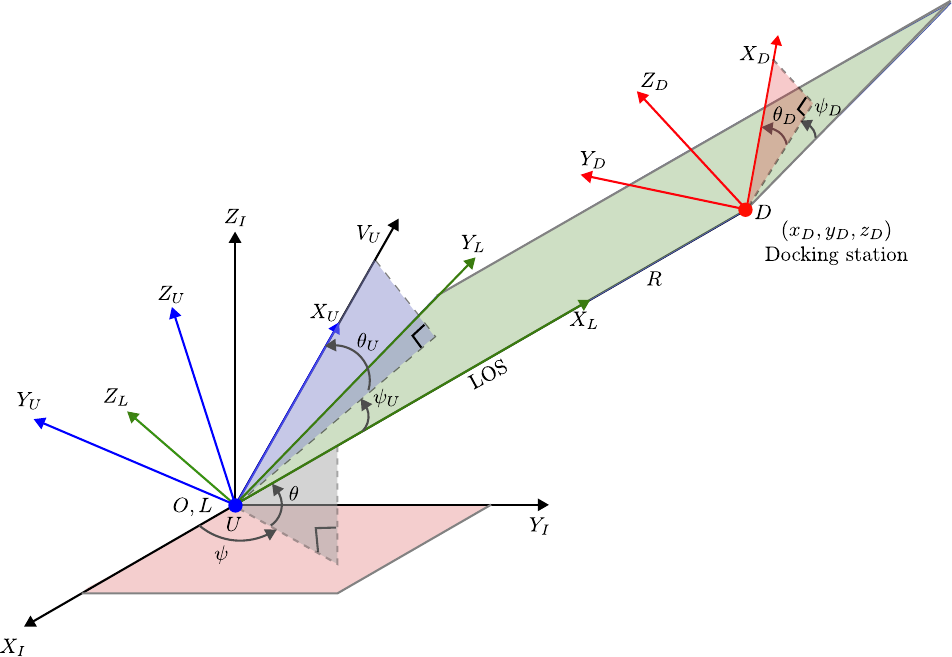}
		\caption{3D engagement scenario of a vehicle with a stationary DS.}
		\label{fig:engagement_stationary_DS}
	\end{figure}
	\Cref{fig:engagement_stationary_DS} shows the representation of the geometric engagement between the current location and orientation of the AV and the DS, along with the coordinate reference frames considered to describe the motion of the AV relative to DS. The inertial earth-fixed coordinate reference frame ($I$-frame) is denoted by $OX_IY_IZ_I$, whose $x,y,z$ axes point towards $OX_I, OY_I$, and $OZ_I$, respectively. 
	The second one is the U-frame, denoted as $UX_UY_UZ_U$, whose $x$-axis is aligned with the AV's velocity vector, and the origin is fixed to the center of gravity of AV.
	The AV is located at $U$, and DS is located at $D$ (see \Cref{fig:engagement_stationary_DS}). The distance between the AV and the DS is denoted by $R$. For brevity of representation, the origins of the $I$-frame and the $U$-frame are shown to be collocated. The third coordinate frame is the L-frame, denoted as $LX_LY_LZ_L$, whose $x$-axis is along the LOS joining $U$ and $D$, and is pointing from $U$ to $D$. Another coordinate frame is the $D$-frame, labeled as $DX_DY_DZ_D$, whose $x$-axis is aligned with the longitudinal reference axis of the body of the DS. 
	Rotating the $I$-frame with azimuth angle, $\psi$, about the $OZ_I$ and the elevation angle, $-\theta$, about the $y$ axes of the rotated $OX_IY_IZ_I$ frame, respectively, aligns $OX_IY_IZ_I$ with $LX_LY_LZ_L$. Similarly, the $LX_LY_LZ_L$ can be aligned with $UX_UY_UZ_U$ by performing two intrinsic rotations by $\psi_U$ and $-\theta_U$ about the $z$ and $y$ axes, respectively. By the same token, rotating by $\psi_D$ and $-\theta_D$ about the $z$ and $y$ axes, respectively, aligns the $LX_LY_LZ_L$ to $DX_DY_DZ_D$.  With the reference frames in place, we now define the orientation at which the AV should approach the DS, as per the available opening port.
	\begin{definition}[Approach angle]
		\label{defn:approach_angle}
		Approach angle is the angle made by the velocity vector of the AV with the velocity vector of the DS. It is defined as a tuple $(\psi_{\text{app}},~\theta_{\text{app}})$, which denotes the relative orientation of $U$-frame with $D$-frame. When $D-$frame is given a rotation by $\psi_{\text{app}}$ about its $z$-axis, followed by another rotation by $-\theta_{\text{app}}$ about the rotated $y$-axes, $D$-frame aligns with $U-$frame. 
	\end{definition}
	\begin{figure}[!h]
		\centering
		\includegraphics[width=0.7\linewidth]{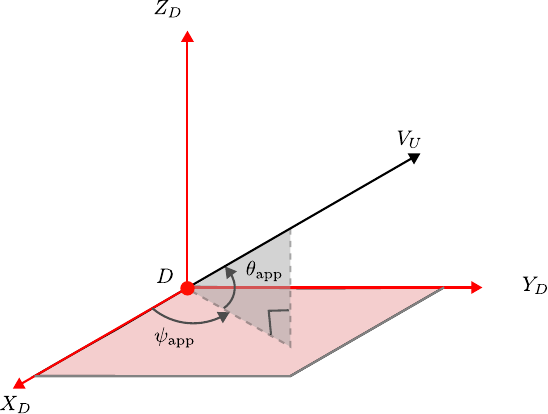}
		\caption{Approach angle of the vehicle to the docking station.}
		\label{fig:approach_angle_stationary}
	\end{figure}
	\begin{problem}
		Consider the engagement scenario as represented in \cref{fig:engagement_stationary_DS}, the objective is to guide the AV to the docking station such that it approaches the DS at a prespecified approach angle while simultaneously controlling its speed to near zero at the time of docking. Mathematically, the docking can be achieved by ensuring that $V_U$ is along $\psiapp$ and $-\thetaapp$ (refer \Cref{fig:approach_angle_stationary}) with respect to the coordinate frame attached to DS ($DX_DY_DZ_D$) and $R$ and $\dot{R}$ go to zero simultaneously.
	\end{problem}
	
	\section{Main Results}\label{sec:main results}
	In this section, we derive a control strategy to achieve autonomous docking with a stationary docking station in three-dimensional space. To address this problem, we see it as a relative motion between the AV and the docking station. 
	To achieve a successful docking, two objectives must be met. One is to approach the docking station at a specified orientation (see \cref{defn:approach_angle}), and the second is to reduce the vehicle's speed to near zero at the time of docking. Both objectives are critical for the mission's success. The proposed strategy achieves the specified orientation to the docking station by using the LOS angles (azimuth and elevation angles) as depicted in \cref{fig:engagement_stationary_DS}. To achieve near-zero speed while approaching the DS, the proposed strategy simultaneously controls AV's speed based on its distance from the DS.
	To this end, we first establish the equations governing the motion of the AV relative to the DS.  
	
	\subsection{Kinematics of 3D engagement}
	To describe the motion of the AV relative to DS, we use polar coordinates and three reference frames: the $I$-, $U$-, $L$-, and $D$-frames as denoted in \cref{fig:engagement_stationary_DS}. The governing equations of motion of the AV is given by
	\begin{align}
		\dot{R}&=-V_U\cos{\theta_U}\cos{\psi_U}, \label{eqn:Rdot} \\
		R\dot{\theta}&=-V_U\sin{\theta_U},\label{eqn:theta_dot} \\
		R\cos\theta\dot{\psi}&=-V_U\cos\theta_U\sin\psi_U,\label{eqn:psi_dot}\\ 
		\dot{\theta}_U&=\dfrac{a_{U_z}}{V_U}-\dot{\psi}\sin\theta\sin\psi_U-\dot{\theta}\cos\psi_U,\label{eqn:thetaU_dot}\\
		\label{eqn:psiU_dot} \dot{\psi}_U&=\dfrac{a_{U_y}}{V_U\cos\theta_U}+\dot{\psi}\tan\theta_U\cos\psi_U\sin\theta -\dot{\psi}\cos\theta \nonumber \\
		& \quad-\dot{\theta}\tan\theta_U\sin\psi_U, \\
		a_{U_x} &= \dot{V}_U. \label{eqn:VU_dot}
	\end{align}
	Here, $V_U$ denotes the speed of the AV, which is controlled using the control input $a_{U_x}$. To change the orientation of the velocity vector of the AV in the yaw and pitch plane, $a_{U_y} $ and $ a_{U_z}$ are the control inputs. Collectively, we denote these control acceleration components of the AV in the U-frame as $\mathcal{U} = [a_{U_x} ~~ a_{U_y} ~~ a_{U_z}]^\top$.
	
	\subsection{Derivation of the docking strategy}
	In this subsection, we derive the guidance and control strategy to simultaneously control the orientation and the speed of the AV to achieve successful docking of the AV with the DS. For successful docking, the AV is commanded to approach the DS at a prespecified orientation depending on the opening port available at the DS. To guide the AV towards the DS, we first obtain the conditions for the two to be on a collision course (that is, the AV's velocity vector is oriented at the specified approach angle and is moving directly towards the stationary DS). In other words, $\dot{R} <0$, $\dot{\theta}=0$, and $\dot{\psi}=0$. Using these conditions in \Cref{eqn:theta_dot,eqn:psi_dot}, we get $\theta_U=0$ and $\psi_U=0$. This indicates that the velocity vector of the AV is aligned with the LOS (line joining AV and the stationary DS). Let the final LOS angles (azimuth and elevation) at the time of docking be denoted by $\psi_F$ and $\theta_F$. Since the DS is stationary, the desired final LOS angles would be the same as the prespecified approach angles. Therefore, the desired orientation for docking can be achieved by driving the LOS angles as follows:
	\begin{align}
		\theta = \theta_F= \thetaapp  ,~ \dot{\theta}=0,~ \psi = \psi_F=\psiapp,~ \dot{\psi}=0 \label{eqn:theta_psi_app}
	\end{align}
	Furthermore, to dock safely, it is crucial to reduce the speed of the AV to near zero, which can be achieved by ensuring that as $R$ approaches zero, $\dot{R}$ also goes to zero. Before proceeding to the control design, let us obtain the dynamics of range ($R$) and LOS angles ($\theta$ and $\psi$). To obtain the dynamics of $R$, on differentiating \cref{eqn:Rdot} with respect to time, we get
	\begin{align}
		\ddot{R} &= - \left( \dot{V}_U\cos\theta_U\cos\psi_U - V_U\dot{\theta}_U\sin\theta_U\cos\psi_U \right. \nonumber \\
		& \quad ~\left. - V_U\dot{\psi}_U\cos\theta_U\sin\psi_U \right). \label{eqn:R_ddot}
	\end{align}
	By substituting $\dot{\theta}_U$ and $\dot{\psi}_U$ from \Cref{eqn:thetaU_dot,eqn:psiU_dot} into \cref{eqn:R_ddot}, we get
	\begin{align}
		\ddot{R} &=  -\dot{V}_U\cos\theta_U\cos\psi_U \nonumber \\ 
		&+ \sin\theta_U\cos\psi_U\left(a_{U_z} - V_U\dot{\psi}\sin\theta\sin\psi_U - V_U\dot{\theta}\cos\psi_U\right)  \nonumber \\ & \quad  +\cos\theta_U\sin\psi_U\left(\dfrac{a_{U_y}}{\cos\theta_U} + V_U\dot{\psi}\tan\theta_U\cos\psi_U\sin\theta \right. \nonumber \\ 
		& \left. \quad - V_U\dot{\psi}\cos\theta - V_U\dot{\theta}\tan\theta_U\sin\psi_U\right). \label{eqn:R_ddot1}
	\end{align}
	Further simplifications of \cref{eqn:R_ddot1} leads to
	\begin{align}
		\ddot{R} &= - \left( \dot{V}_U\cos\theta_U\cos\psi_U - a_{U_z}\sin\theta_U\cos\psi_U \right. \nonumber \\
		& \left. - a_{U_y}\sin\psi_U + V_U\dot{\theta}\sin\theta_U + V_U\dot{\psi}\cos\theta_U\sin\psi_U\cos\theta \right) \label{eqn:R_ddot2}
	\end{align}
	Now, for convenience \cref{eqn:R_ddot2} can be rewritten compactly as
	\begin{align}
		\ddot{R} = f_R + g_{R_1}\dot{V}_U + g_{R_2}a_{U_y} + g_{R_3}a_{U_z},\label{eqn:R_ddot_compact}
	\end{align}
	where the terms $f_R,g_{R_1},g_{R_2}$, and $g_{R_3}$ are defined as 
	\begin{align*}
		f_R &= - V_U\dot{\theta}\sin\theta_U - V_U\dot{\psi}\cos\theta_U\sin\psi_U\cos\theta, \\
		g_{R_1} &= - \cos\theta_U\cos\psi_U,
		g_{R_2} = \sin\psi_U,
		g_{R_3} = \sin\theta_U\cos\psi_U.
	\end{align*} 
	On differentiating \cref{eqn:theta_dot} with respect to time, we get
	\begin{align}
		\dot{R}\dot{\theta} + R\ddot{\theta} =  - (\dot{V}_U\sin\theta_U + V_U\dot{\theta}_U\cos\theta_U). \label{eqn:theta_ddot1}
	\end{align}
	Using \cref{eqn:thetaU_dot} into \cref{eqn:theta_ddot1}, we get
	\begin{align}
		\dot{R}\dot{\theta} + R\ddot{\theta} &=  -\dot{V}_U\sin\theta_U - \cos\theta_U\left(a_{U_z} - V_U\dot{\psi}\sin\theta\sin\psi_U \right. \nonumber \\ 
		& \left. \quad - V_U\dot{\theta}\cos\psi_U\right)\label{eqn:theta_ddot2}
	\end{align}
	Now, \cref{eqn:theta_ddot2}, can be rewritten as
	\begin{align}
		\ddot{\theta} &= \dfrac{1}{R} \left[  -\dot{V}_U\sin\theta_U - a_{U_z}\cos\theta_U + V_U\dot{\psi}\cos\theta_U\sin\theta\sin\psi_U \right. \nonumber \\ & \left. \quad + V_U\dot{\theta}\cos\theta_U\cos\psi_U  - \dot{R}\dot{\theta} \right],
	\end{align}
	which can be further expressed for convenience as
	\begin{align}
		\ddot{\theta} = f_\theta + g_{\theta_1}\dot{V}_U + g_{\theta_2}a_{U_y} + g_{\theta_3}a_{U_z},\label{eqn:theta_ddot_compact}
	\end{align}
	where the terms $f_\theta,g_{\theta_1},g_{\theta_2}$, and $g_{\theta_3}$ are defined as 
	\begin{align*}
		f_\theta &= \dfrac{1}{R} \left[  V_U\dot{\psi}\cos\theta_U\sin\theta\sin\psi_U + V_U\dot{\theta}\cos\theta_U\cos\psi_U - \dot{R}\dot{\theta} \right], \\
		g_{\theta_1} &= -\sin\theta_U/R,~~
		g_{\theta_2} = 0,~~
		g_{\theta_3} = -\cos\theta_U/R.
	\end{align*} 
	Next, on differentiating \cref{eqn:psi_dot} with respect to time gives,
	\begin{align}
		\dot{R}&\cos\theta\dot{\psi} - R\sin\theta\dot{\theta}\dot{\psi} + R\cos\theta\ddot{\psi} = -\dot{V}_U\cos\theta_U\sin\psi_U \nonumber \\
		& + V_U\dot{\theta}_U\sin\theta_U\sin\psi_U - V_U\dot{\psi}_U\cos\theta_U\cos\psi_U . \label{eqn:psi_ddot3}
	\end{align}
	Substituting $\dot{\theta}_U$ and $\dot{\psi}_U$ from \Cref{eqn:thetaU_dot,eqn:psiU_dot} into \cref{eqn:psi_ddot3}, we get
	\begin{align}
		\dot{R}&\cos\theta\dot{\psi} - R\sin\theta\dot{\theta}\dot{\psi} + R\cos\theta\ddot{\psi} =  -\dot{V}_U\cos\theta_U\sin\psi_U 
		\nonumber \\ 
		&  + \sin\theta_U\sin\psi_U\left(a_{U_z} - V_U\dot{\psi}\sin\theta\sin\psi_U - V_U\dot{\theta}\cos\psi_U\right)  \nonumber \\ 
		&  - \cos\theta_U\cos\psi_U\left(\dfrac{a_{U_y}}{\cos\theta_U} + V_U\dot{\psi}\tan\theta_U\cos\psi_U\sin\theta \right. \nonumber \\ 
		& \left. - V_U\dot{\psi}\cos\theta - V_U\dot{\theta}\tan\theta_U\sin\psi_U\right). \label{eqn:psi_ddot1}
	\end{align}
	Simplifying \cref{eqn:psi_ddot1}, we can obtain
	\begin{align}
		\ddot{\psi} &= \dfrac{1}{R\cos\theta} \left[ -\dot{V}_U\cos\theta_U\sin\psi_U + a_{U_z}\sin\theta_U\sin\psi_U \right. \nonumber \\ 
		& \left. \quad - a_{U_y}\cos\psi_U - V_U\dot{\psi}\sin\theta_U\sin\theta - \dot{R}\cos\theta\dot{\psi} \right. \nonumber \\ & \quad \left. + V_U\dot{\psi}\cos\theta_U\cos\psi_U\cos\theta + R\sin\theta\dot{\theta}\dot{\psi} \right]\label{eqn:psi_ddot2}
	\end{align}
	Now, \cref{eqn:psi_ddot2} can be written in a convenient form as
	\begin{align}
		\ddot{\psi} = f_\psi + g_{\psi_1}\dot{V}_U + g_{\psi_2}a_{U_y} + g_{\psi_3}a_{U_z},\label{eqn:psi_ddot_compact}
	\end{align}
	where the terms $f_\psi,g_{\psi_1},g_{\psi_2}$, and $g_{\psi_3}$ are defined as 
	\begin{align*}
		f_\psi &= \dfrac{1}{R\cos\theta} \left[  - V_U\dot{\psi}\sin\theta_U\sin\theta + V_U\dot{\psi}\cos\theta_U\cos\psi_U\cos\theta \right. \nonumber \\ & \quad \left. 
		- \dot{R}\cos\theta\dot{\psi} + R\sin\theta\dot{\theta}\dot{\psi} \right], \\
		g_{\psi_1} &= \dfrac{-\cos\theta_U\sin\psi_U}{R\cos\theta}, 
		g_{\psi_2} = \dfrac{-\cos\psi_U}{R\cos\theta}, 
		g_{\psi_3} = \dfrac{\sin\theta_U\sin\psi_U}{R\cos\theta}.
	\end{align*}
	With the range and LOS error dynamics in place, we now proceed to the controller design. Note that the range and LOS error dynamics as given by \cref{eqn:R_ddot_compact,eqn:theta_ddot_compact,eqn:psi_ddot_compact}, have a relative degree of two with respect to the available control inputs.
	
	\subsection{Controller design}
	Next, to proceed with the controller design, let us define the error in LOS angles using \cref{eqn:theta_psi_app} as $e_\theta = \theta - \theta_F$, $e_\psi = \psi - \psi_F$. As the dynamics of range and LOS angles have a relative degree of two, the sliding surfaces are chosen as  
	\begin{align}
		\mathcal{S} = \begin{bmatrix}
			\mathcal{S}_R \\ \mathcal{S}_\theta \\ \mathcal{S}_\psi
		\end{bmatrix} = \begin{bmatrix}
			\dot{R} + k_R R \\
			\dot{e}_\theta + k_\theta e_\theta \\
			\dot{e}_\psi + k_\psi e_\psi 
		\end{bmatrix}=
		\begin{bmatrix}
			\dot{R} + k_R R \\
			\dot{\theta} + k_\theta(\theta - \theta_F) \\
			\dot{\psi} + k_\psi (\psi -\psi_F)
		\end{bmatrix}.\label{eqn:S_vec}
	\end{align}
	Consider a quadratic Lyapunov function candidate given by $\mathcal{W} = 0.5\mathcal{S}^\top \mathcal{S}$,
	whose time derivative can be obtained as 
	\begin{align}
		\dot{\mathcal{W}} = \mathcal{S}^\top \dot{\mathcal{S}}.\label{eqn:Lyapunov_dot}
	\end{align}
	Now, to compute the derivative of sliding surfaces, differentiating \cref{eqn:S_vec} with time yields
	\begin{align}
		\dot{\mathcal{S}} = \begin{bmatrix}
			\dot{\mathcal{S}}_R \\ \dot{\mathcal{S}}_\theta \\ \dot{\mathcal{S}}_\psi
		\end{bmatrix} = \begin{bmatrix}
			\ddot{R} + k_R \dot{R} \\
			\ddot{\theta} + k_\theta\dot{\theta} \\
			\ddot{\psi} + k_\psi \dot{\psi}
		\end{bmatrix}\label{eqn:S_dot}
	\end{align}
	Using \cref{eqn:R_ddot_compact,eqn:theta_ddot_compact,eqn:psi_ddot_compact} into \cref{eqn:S_dot} results in 
	\begin{align}
		\dot{\mathcal{S}} = \begin{bmatrix}
			f_R + g_{R_1}\dot{V}_U + g_{R_2}a_{U_y} + g_{R_3}a_{U_z} + k_R \dot{R} \\
			f_\theta + g_{\theta_1}\dot{V}_U + g_{\theta_2}a_{U_y} + g_{\theta_3}a_{U_z} + k_\theta\dot{\theta} \\
			f_\psi + g_{\psi_1}\dot{V}_U + g_{\psi_2}a_{U_y} + g_{\psi_3}a_{U_z} + k_\psi \dot{\psi}
		\end{bmatrix}. \label{eqn:S_dot1}
	\end{align}
	Equation \cref{eqn:S_dot1} can be rewritten in matrix-vector form as
	\begin{align}
		\dot{\mathcal{S}} = \mathcal{F} + \mathcal{G} \mathcal{U}, \label{eqn:S_dot_matrix}
	\end{align}
	where the terms $\mathcal{F}$ and $\mathcal{G}$ are defined as 
	\begin{align}
		\mathcal{F} = \begin{bmatrix}
			f_1 \\ f_2 \\ f_3
		\end{bmatrix}=\begin{bmatrix}
			f_R +k_R \dot{R}  \\ f_\theta + k_\theta\dot{\theta} \\ f_\psi + k_\psi \dot{\psi}
		\end{bmatrix}, \mathcal{G} = \begin{bmatrix}
			g_{R_1} & g_{R_2} & g_{R_3} \\
			g_{\theta_1} & g_{\theta_2} & g_{\theta_3} \\
			g_{\psi_1} & g_{\psi_2} & g_{\psi_3}
		\end{bmatrix}. \label{eqn:FG}
	\end{align}
	\begin{remark}[Invertibility of $\mathcal{G}$]
		Alternatively, the inverse of $\mathcal{G}$ can also be computed by rewriting it as
		\begin{align*}
			\mathcal{G} &= \begin{bmatrix}  -\cos\theta_U\cos\psi_U & \sin\psi_U & \sin\theta_U\cos\psi_U \\  -\frac{\sin\theta_U}{R} & 0 & -\frac{\cos\theta_U}{R} \\  -\frac{\cos\theta_U\sin\psi_U}{R\cos\theta} & -\frac{\cos\psi_U}{R\cos\theta} & \frac{\sin\theta_U\sin\psi_U}{R\cos\theta}   \end{bmatrix} = \mathcal{G}_R \mathcal{G}_\theta,
		\end{align*}
		where $\mathcal{G}_R$ and $\mathcal{G}_\theta$ are taken as
		\begin{align}
			\mathcal{G}_R&= \begin{bmatrix}  1 & 0 & 0 \\  0 & \frac{1}{R} & 0 \\  0 & 0 & \frac{1}{R\cos\theta}  \end{bmatrix},  \nonumber \\
			\mathcal{G}_\theta &= \begin{bmatrix}  -\cos\theta_U\cos\psi_U & \sin\psi_U & \sin\theta_U\cos\psi_U \\  -\sin\theta_U & 0 & -\cos\theta_U \\  -\cos\theta_U\sin\psi_U & -\cos\psi_U & \sin\theta_U\sin\psi_U  \end{bmatrix}. \label{eqn:GR_Gtheta}
		\end{align}
		By using the property of matrix inverse, $\mathcal{G}^{-1}$ can be computed as 
		\begin{align}
			\mathcal{G}^{-1} = \left(\mathcal{G}_R \mathcal{G}_\theta\right)^{-1} = \mathcal{G}_\theta^{-1} \mathcal{G}_R^{-1}. \label{eqn:Ginv1}
		\end{align}
		From \Cref{eqn:GR_Gtheta}, one can observe and verify that $\mathcal{G}_R$ is a diagonal matrix, while $\mathcal{G}_\theta$ is an orthogonal matrix. The inverse of a diagonal matrix can be computed by taking the reciprocal of its diagonal elements, whereas the inverse of the orthogonal matrix is simply its transpose. By using these matrix properties of diagonal matrix and orthogonal matrix into \Cref{eqn:Ginv1}, the inverse of $\mathcal{G}$ can be computed as $\mathcal{G}^{-1}$
		\begin{align}
			=\begin{bmatrix}  -\cos\theta_U\cos\psi_U & -R\sin\theta_U & -R\cos\theta\cos\theta_U\sin\psi_U \\  \sin\psi_U & 0 & -R\cos\theta\cos\psi_U \\  \sin\theta_U\cos\psi_U & -R\cos\theta_U & R\cos\theta\sin\theta_U\sin\psi_U  \end{bmatrix}. \label{eqn:Ginv_final}
		\end{align}
	\end{remark}
	By choosing the control input $\mathcal{U}$ as
	\begin{equation}
		\mathcal{U} = -\mathcal{G}^{-1} \mathcal{F} - \mathcal{G}^{-1}\begin{bmatrix}
			\tilde{M}_R|\mathcal{S}_R|^\alpha\sign(\mathcal{S}_R) + N_R \mathcal{S}_R \\
			\tilde{M}_\theta|\mathcal{S}_{\theta}|^\alpha\sign(\mathcal{S}_{\theta}) + N_\theta \mathcal{S}_{\theta}\\
			\tilde{M}_\psi|\mathcal{S}_{\psi}|^\alpha\sign(\mathcal{S}_{\psi}) + N_\psi \mathcal{S}_{\psi}
		\end{bmatrix}, \label{eqn:control_accn}
	\end{equation}
	and substituting \cref{eqn:control_accn} into \Cref{eqn:Lyapunov_dot}, one can obtain
	\begin{align}
		\nonumber \dot{\mathcal{W}} &= \begin{bmatrix}
			\mathcal{S}_R & \mathcal{S}_\theta & \mathcal{S}_\psi
		\end{bmatrix} \begin{bmatrix}
			-\tilde{M}_R|\mathcal{S}_R|^\alpha\sign(\mathcal{S}_R) - N_R \mathcal{S}_R \\
			-\tilde{M}_\theta|\mathcal{S}_{\theta}|^\alpha\sign(\mathcal{S}_{\theta}) - N_\theta \mathcal{S}_{\theta}\\
			-\tilde{M}_\psi|\mathcal{S}_{\psi}|^\alpha\sign(\mathcal{S}_{\psi}) - N_\psi \mathcal{S}_{\psi}
		\end{bmatrix} \\ \nonumber
		&=  \mathcal{S}_R \left( -\tilde{M}_R|\mathcal{S}_R|^\alpha\sign(\mathcal{S}_R) - N_R \mathcal{S}_R\right) \\ \nonumber
		& \quad +\mathcal{S}_{\theta} \left(-\tilde{M}_\theta|\mathcal{S}_{\theta}|^\alpha\sign(\mathcal{S}_{\theta}) - N_\theta \mathcal{S}_{\theta} \right)  \\
		& \quad + \mathcal{S}_\psi \left(-\tilde{M}_\psi|\mathcal{S}_{\psi}|^\alpha\sign(\mathcal{S}_{\psi}) - N_\psi \mathcal{S}_{\psi}\right) \nonumber \\
		&= -\sum_i \left( \tilde{M}_i|\mathcal{S}_i|^{\alpha+1} +N_i \mathcal{S}_i^2 \right), \forall~ i \in \{R,~\theta,~\psi \},
		\label{eqn:Wdot}
	\end{align}
	where $0<\alpha<1$ and the controller parameters $\tilde{M}_R$, $N_R$,  $\tilde{M}_\theta$, $N_\theta$, $\tilde{M}_\psi$, and $N_\psi$ are all positive real numbers.
	Now, \Cref{eqn:Wdot} can be rewritten in the form of an inequality as
	\begin{align}
		\dot{\mathcal{W}} &\leq -\min_i(\tilde{M}_i) \sum_i \left( \mathcal{S}_i^2 \right)^{\frac{\alpha+1}{2}} - \min_i(N_i) \sum_i \mathcal{S}_i^2 \nonumber \\ 
		&\leq -k_1 \mathcal{W} - k_2 \mathcal{W}^\gamma \leq-(k_1\mathcal{W}^{1-\gamma}+k_2)\mathcal{W}^\gamma, \label{eqn:Wdot_ineq}
	\end{align}
	where the constants are defined as $k_1 = 2\min_i(N_i) > 0$, $k_2 = 2^{\frac{\alpha+1}{2}}\min_i(\tilde{M}_i) > 0$, and $\gamma = \frac{\alpha+1}{2}$.
	By separating the variables in \Cref{eqn:Wdot_ineq} and using $v=\mathcal{W}^{1-\gamma}$ gives
	$$\frac{1}{1-\gamma}\frac{dv}{k_1v+k_2}\leq-dt,$$
	which on integrating from the initial time $t=0$ (where $\mathcal{W}=\mathcal{W}(0)$) to the settling time $t=T$ (where $\mathcal{W}(T)=0$), yields 
	\begin{align}
		T\leq\frac{1}{k_1(1-\gamma)}\ln\left(1+\frac{k_1}{k_2}\mathcal{W}^{1-\gamma}(0)\right). \label{eqn:T}
	\end{align}
	It follows from \Cref{eqn:T} that the system reaches the equilibrium $\mathcal{W}=0$ (that is, $\mathcal{S}$ converges to zero) in a bounded, finite time $T$, which depends directly on the chosen controller gains $k_1$, $k_2$, the exponent $\gamma$, and the initial conditions.
	Once the sliding mode is enforced on the chosen sliding surfaces, the reduced-order dynamics are given by 
	$\dot{z} = -k_iz$ for $i\in\{R,\theta,\psi\}$ and $z\in \{R,e_\theta, e_\psi\}$. On integrating, we obtain $z(t) = z(0)e^{-k_it}.$ Thus, $R\rightarrow 0$, $\theta \rightarrow \theta_F$, and $\psi \rightarrow \psi_F$ as $t\rightarrow \infty$.
	
	\subsection{Analysis of behavior of control law when AV is nearby DS}
	From the derived control law in \Cref{eqn:control_accn}, it is evident that as the AV nears the DS at the desired orientation, the second term goes to zero. This is because the sliding surfaces approach zero. Now, for the first term, we need to have a closer look at the expressions for the elements of $f_R$, $f_\theta$, and $f_\psi$. Using \Cref{eqn:Rdot,eqn:theta_dot} into $f_R$, we get
	\begin{align}
		f_R &= - V_U \left(-\frac{V_U\sin\theta_U}{R}\right)\sin\theta_U \nonumber \\
		& \quad - V_U \left(-\frac{V_U\cos\theta_U\sin\psi_U}{R\cos\theta}\right)\cos\theta_U\sin\psi_U\cos\theta \nonumber \\
		&= \frac{V_U^2}{R} \left( \sin^2\theta_U + \cos^2\theta_U \sin^2\psi_U \right).
		\label{eqn:fR_final}
	\end{align}
	Further, using \Cref{eqn:Rdot,eqn:theta_dot,eqn:psi_dot} into $f_\theta$, it can be written as
	\begin{align*}
		f_\theta &= \dfrac{V_U}{R} \left(-\frac{V_U\cos\theta_U\sin\psi_U}{R\cos\theta}\right)\cos\theta_U\sin\theta\sin\psi_U \nonumber \\
		& \quad + \dfrac{V_U}{R} \left(-\frac{V_U\sin\theta_U}{R}\right)\cos\theta_U\cos\psi_U \nonumber \\
		& - \dfrac{1}{R}(-V_U\cos\theta_U\cos\psi_U) \left(-\frac{V_U\sin\theta_U}{R}\right), 
	\end{align*}
	which can be simplified to
	\begin{align}
		f_\theta = -\frac{V_U^2}{R^2} \left( \cos^2\theta_U\sin^2\psi_U\tan\theta + 2\sin\theta_U\cos\theta_U\cos\psi_U \right). \label{eqn:ftheta_final}
	\end{align}
	Similarly, we obtain $f_\psi$ using \Cref{eqn:theta_dot,eqn:psi_dot} as
	\begin{align}
		f_\psi = \dfrac{2 V_U^2}{R^2 \cos\theta} \sin\psi_U \cos\theta_U \left( \sin\theta_U \tan\theta - \cos\theta_U \cos\psi_U \right). \label{eqn:fpsi_final}
	\end{align}
	Using \Cref{eqn:fR_final,eqn:ftheta_final,eqn:fpsi_final,eqn:Ginv_final}, we can see that $\mathcal{F}$ contains $R$ in the denominator but also contain $\sin \theta_U$ or $\sin \psi_U$ or both in the numerator. The AV first aligns itself at the desired orientation before $R \rightarrow 0$. This means $\theta_U$ and $\psi_U$ go to zero before $R$ goes to zero. Therefore, one can conclude that the derived control input in \Cref{eqn:control_accn} remains finite even when the AV is about to dock with the DS. 
	
	Now, let us compute the components of the control inputs. Using \Cref{eqn:Rdot,eqn:theta_dot,eqn:psi_dot,eqn:fR_final,eqn:ftheta_final,eqn:fpsi_final} into \Cref{eqn:FG}, we can obtain the elements of $\mathcal{F}$ as
	\begin{subequations}\label{eqn:F123}
		\begin{align}
			f_1 &= \frac{V_U^2}{R} \left( \sin^2\theta_U + \cos^2\theta_U \sin^2\psi_U \right) -k_R V_U\cos\theta_U\cos\psi_U, \\
			f_2 &= -\frac{V_U^2}{R^2} \left( \cos^2\theta_U\sin^2\psi_U\tan\theta + 2\sin\theta_U\cos\theta_U\cos\psi_U \right) \nonumber \\ 
			& \quad -k_\theta \frac{V_U\sin\theta_U}{R}, \\
			f_3 &= \dfrac{2 V_U^2}{R^2 \cos\theta} \sin\psi_U \cos\theta_U \left( \sin\theta_U \tan\theta - \cos\theta_U \cos\psi_U \right) \nonumber \\ 
			& \quad -k_\psi \frac{V_U\cos\theta_U\sin\psi_U}{R\cos\theta}.
		\end{align}
	\end{subequations}
	By using \Cref{eqn:F123,eqn:Ginv_final}, and trigonometric identity $\sin^2\theta_U + \cos^2\theta_U\sin^2\psi_U = 1 - \cos^2\theta_U\cos^2\psi_U$, along with further simplification, one can compute $\mathcal{G}^{-1}\mathcal{F} = \left[\mathcal{U}_{f1}~\mathcal{U}_{f2}~\mathcal{U}_{f3} \right]^\top$ as
	\begin{subequations}
		\begin{align}
			\mathcal{U}_{f1} &= \frac{V_U^2}{R} \left[ \cos\theta_U\cos\psi_U(1 - \cos^2\theta_U\cos^2\psi_U) \right. \nonumber \\
			&\quad \left. - \sin\theta_U\cos^2\theta_U\sin^2\psi_U\tan\theta \right] \nonumber \\
			&\quad + V_U \left( k_R \cos^2\theta_U\cos^2\psi_U + k_\theta \sin^2\theta_U \right. \nonumber \\
			&\quad \left. + k_\psi \cos^2\theta_U\sin^2\psi_U \right), \label{eqn:Uf1}\\
			\mathcal{U}_{f2} &= \frac{V_U^2}{R} \sin\psi_U \left[ 1 + \cos^2\theta_U\cos^2\psi_U \right. \nonumber \\
			&\quad \left. - 2\sin\theta_U\cos\theta_U\cos\psi_U\tan\theta \right] \nonumber \\
			&\quad + V_U\cos\theta_U\sin\psi_U\cos\psi_U (k_\psi - k_R), \label{eqn:Uf2}\\
			\mathcal{U}_{f3} &= \frac{V_U^2}{R} \left[ \sin\theta_U\cos\psi_U ( 1 + \cos^2\theta_U\cos^2\psi_U ) \right. \nonumber \\
			&\quad \left. + \cos\theta_U\sin^2\psi_U\tan\theta ( 1 + \sin^2\theta_U ) \right] \nonumber \\
			&\quad + V_U\sin\theta_U\cos\theta_U \left( k_\theta - k_R\cos^2\psi_U \right. \nonumber \\
			&\quad \left. - k_\psi\sin^2\psi_U \right). \label{eqn:Uf3}
		\end{align}
	\end{subequations}
	Once the sliding mode is enforced, the reduced order-dynamics comes into effect while $-\mathcal{G}^{-1}\mathcal{F}$ acts on the system. From \Cref{eqn:Uf1,eqn:Uf2,eqn:Uf3}, it can be noted that the expressions have $R$ in their denominators, which raises concerns about the explosion of the control effort requirement as the AV nears the DS. However, AV first aligns itself in the desired orientation; that is, the lead angles $\theta_U$ and $\psi_U$ go to zero before $R \rightarrow 0$. This ensures that \Cref{eqn:Uf1,eqn:Uf2,eqn:Uf3} won't explode as AV nears DS. From \Cref{eqn:Uf1}, it can be noted that $\mathcal{U}_{f1}$ reduces to $k_R V_U$, since $\theta_U \rightarrow 0$ and $\psi_U \rightarrow 0$. Similarly, from \Cref{eqn:Rdot}, it is evident that $\dot{R} = -V_U$. Therefore, $\mathcal{U}_{f1}$ becomes $-k_R\dot{R}$. From \Cref{eqn:S_vec}, the reduced order range dynamics becomes $\dot{R} = -k_R R$. Substituting this into $\mathcal{U}_{f1}$ leads to $\mathcal{U}_{f1} \rightarrow k_R^2 R$. Consequently, the control demand varies linearly with R as the AV nears the DS.
	\section{Simulation}
	In this section, we validate the performance of the proposed docking strategy for an AV in \Cref{eqn:control_accn} using MATLAB\textsuperscript{\textregistered} simulations. The controller parameters used for simulations are as follows: $k_R = 1$, $k_\theta = 0.1$, and $k_\psi = 0.1$, $M_R = 0.0317$, $M_\theta = 0.6963$, $M_\psi = 0.7$, $N_R = 0.0766$, $N_\theta = 0.0178$, and $N_\psi = 0.01$, with $\alpha = 0.9$. These controller parameters are systematically chosen. For example, the gains subscripted with $R$, $\theta$, and $\psi$ are related to range, and LOS angles, respectively. In the simulations, we demonstrate the docking for various initial locations of the AV and the DS. The proposed strategy is also validated for different approach angles to the DS, since the docking port can be oriented in any direction. 
	We represent the initial configurations of the AV as $[x,~y,~z,~V_U,~\psi_U,~\theta_U ]$, where $(x,~y,~z)$ denotes the position (in meters) of the AV in Cartesian coordinates, $V_U$ denote the speed (in m/s) of the AV, ($\psi_U,~\theta_U$) denote the azimuth and elevation angle of the velocity vector of the AV with respect to the LOS joining the AV and the DS. Similarly, we represent the initial configurations of the DS as  $[x,~y,~z,~\psi_F,~\theta_F ]$, where $(x,~y,~z)$ denotes the Cartesian coordinates of DS, ($\psi_F,~\theta_F$) denotes the desired azimuth and elevation angles of the LOS at the time of docking.

	First, we consider a scenario in which the AV is positions are three different initial locations denoted by $P_1: [0,~ 0,~ 0,~ 1.0,~ 10^\circ,~ 20^\circ]$, $P_2: [10,~ 0,~ 0,~ 0.5,~ 30^\circ,~ 60^\circ]$, and $P_3: [0,~ 10,~ 0,~ 1.5,~ 60^\circ,~ 0^\circ]$.  The docking station is located at ($10,~10,~10$) meters with the desired orientation angles as $\psi_F = -45^\circ$ and $\theta_F=-45^\circ$. 
	\begin{figure*}[!t]
		\centering
		\subfloat[Path of the vehicle.]{
			\includegraphics[width=0.23\linewidth]{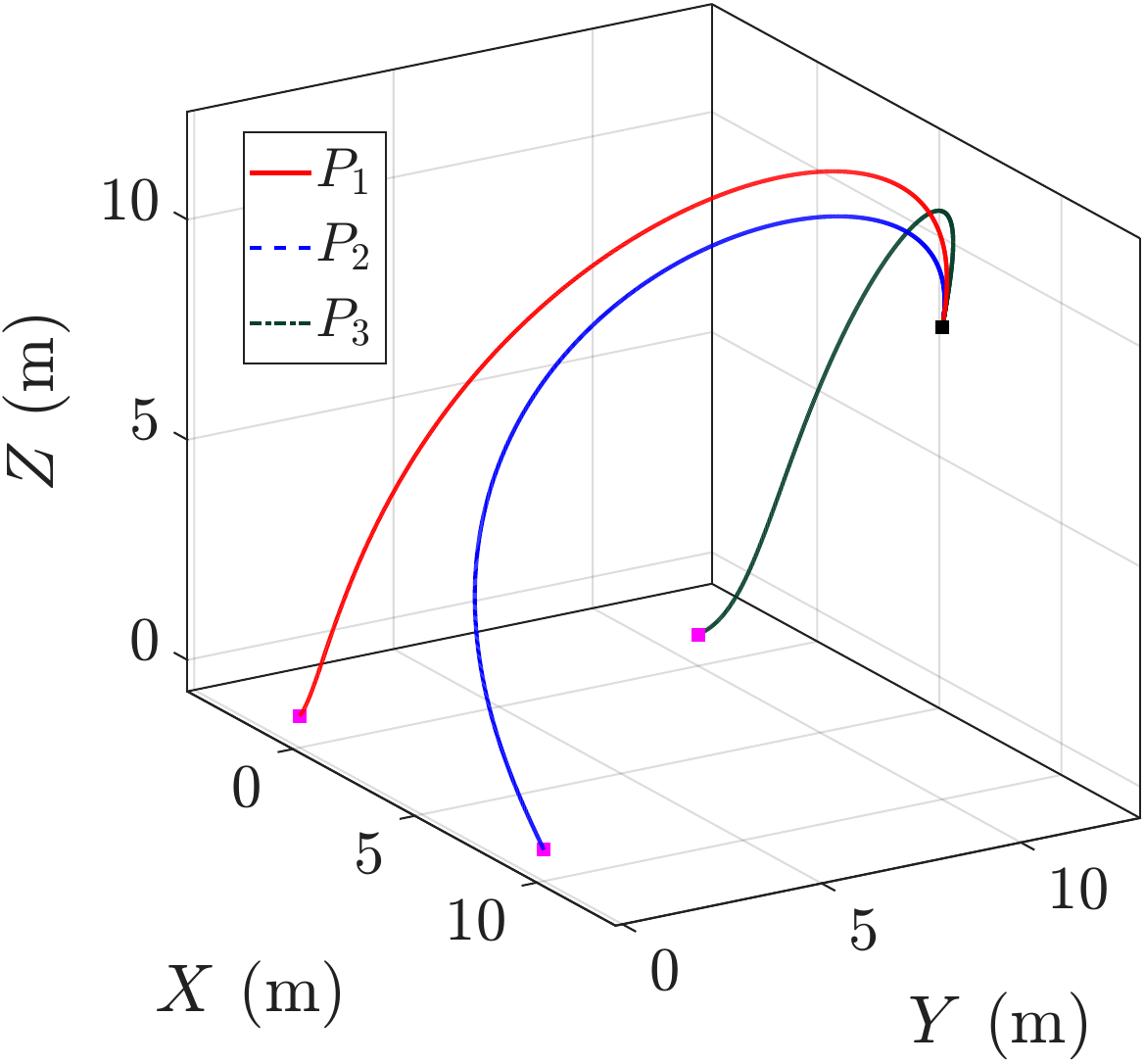}
			\label{fig:P_path}
		}%
		\subfloat[Speed, range and closing rate.]{
			\includegraphics[width=0.23\linewidth]{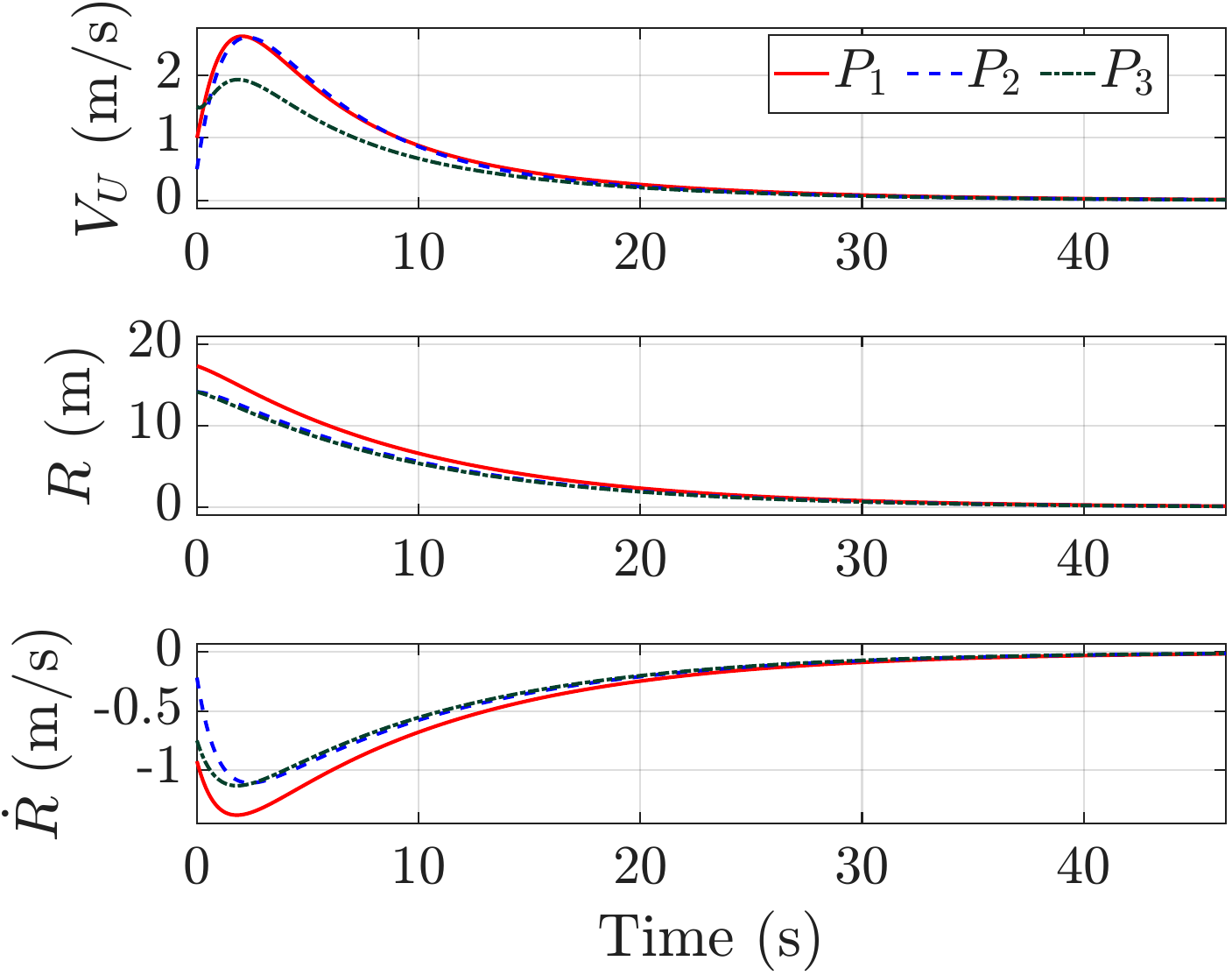}
			\label{fig:P_speed_range}
		}%
		\subfloat[Control acceleration.]{
			\includegraphics[width=0.23\linewidth]{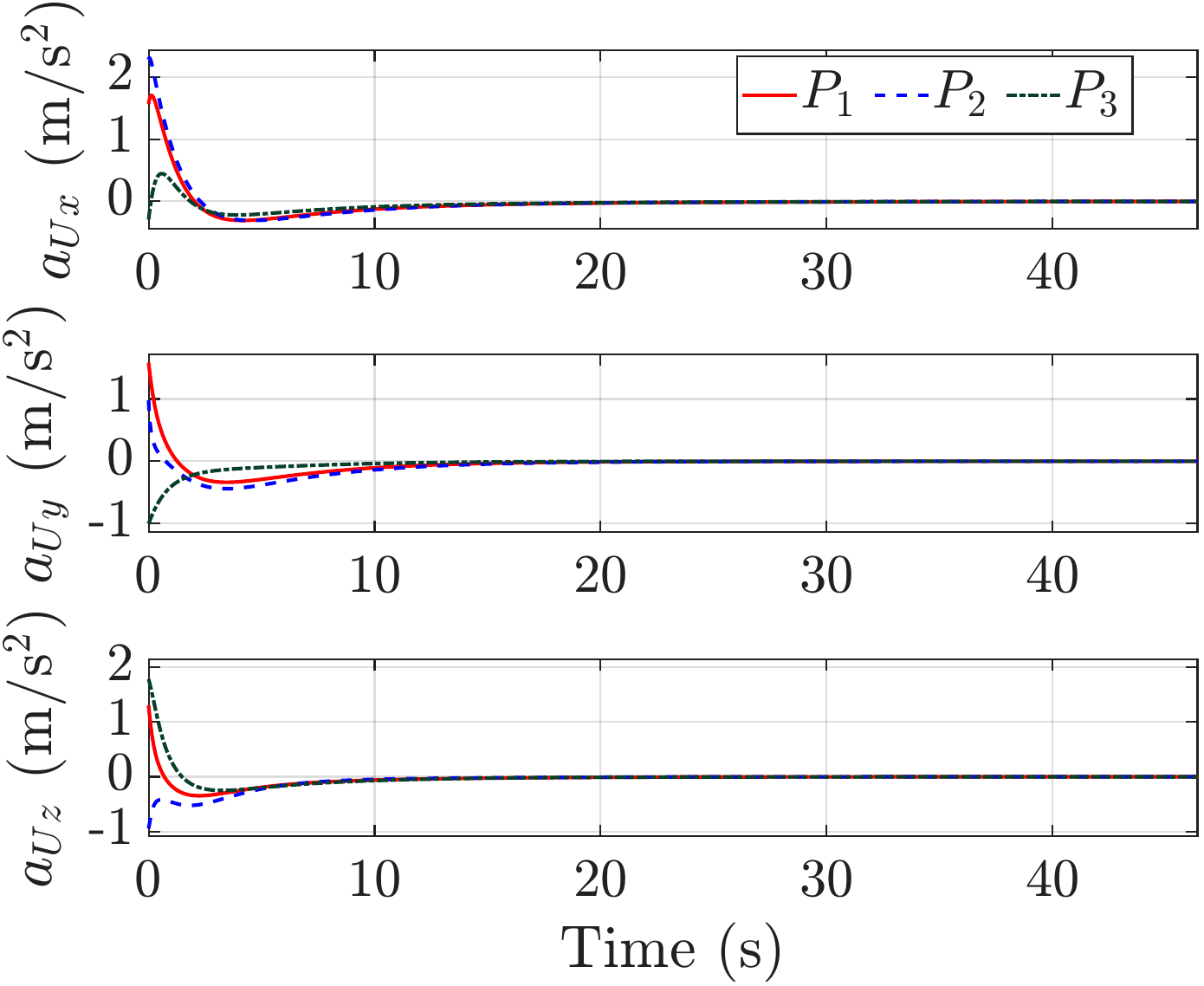}
			\label{fig:P_ctrl_accn}
		}%
		\subfloat[LOS angles and Lyapunov function.]{
			\includegraphics[width=0.23\linewidth]{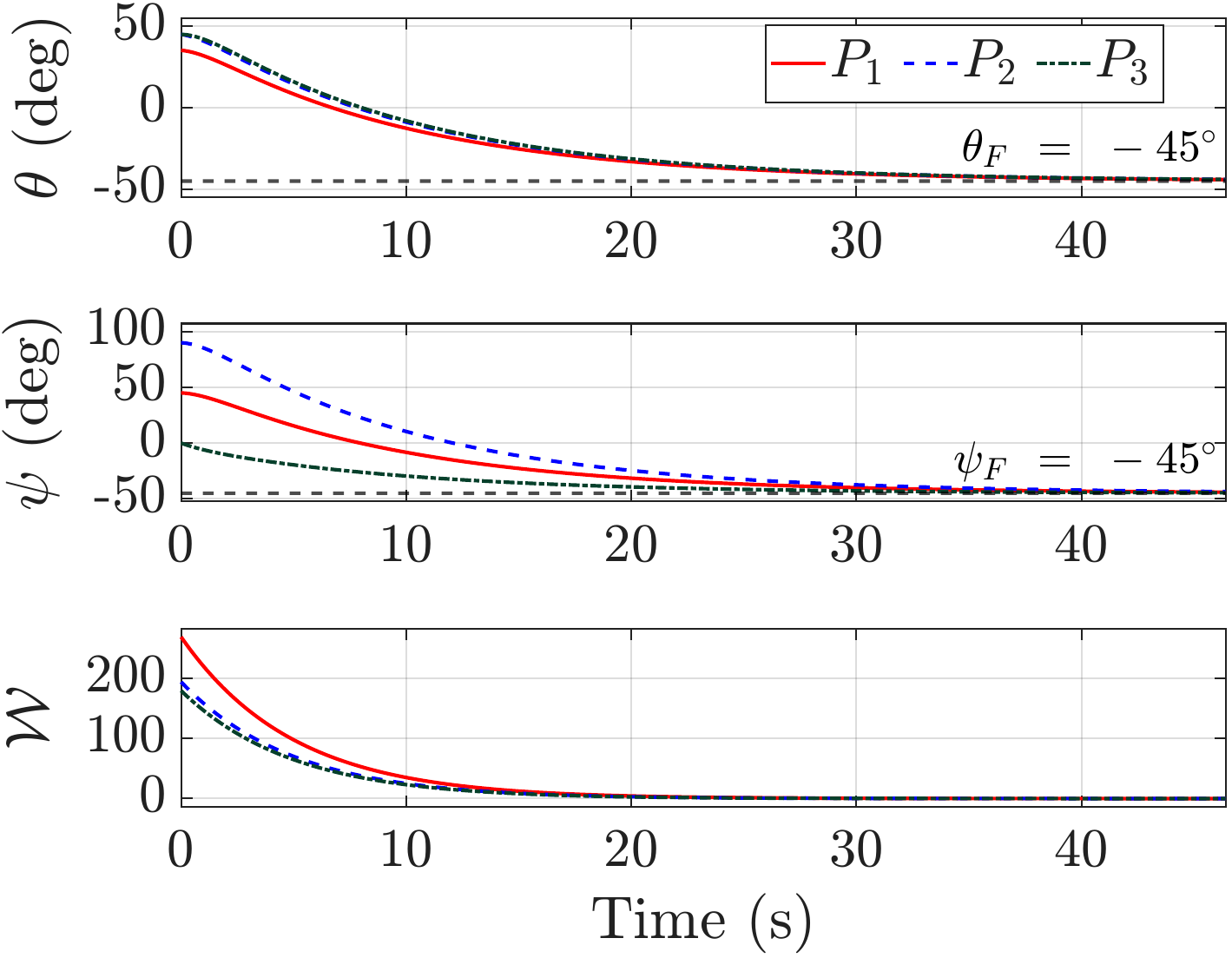}
			\label{fig:P_angles}
		}
		\caption{Demonstration of docking for various initial locations of the vehicle while fixed docking station.}
		\label{fig:P}
	\end{figure*}
	The docking performance for the cases $P_1$, $P_2$, and $P_3$ is demonstrated in \cref{fig:P}. \cref{fig:P_path} shows the path taken by the AV, where the filled magenta squares denote the starting location of the AV and the filled black square denotes the location of the DS. The same notation has been followed for denoting the initial locations of the AV and DS in the latter part of the manuscript unless stated otherwise. 
	To guide itself to the dock station, the AV first increases its speed and, as it nears the DS, reduces its speed to zero (refer \cref{fig:P_speed_range}). This is desirable behavior for docking safely without colliding with the DS. From \cref{fig:P_speed_range} it is evident that as $R\to 0$, $V_U$ also goes to zero. Before the AV attempts to dock, it aligns itself in the specified orientation (see \cref{fig:P_angles}. The Lyapunov function value also decreases to zero as shown in \cref{fig:P_angles}. The control command issued to the AV is depicted in \cref{fig:P_ctrl_accn}. Due to large initial deviations from the desired values, the control demand is initially high, which later reduces as the AV approaches the DS.
	
\begin{figure*}[!t]
		\centering
		\subfloat[Path of the vehicle.]{
			\includegraphics[width=0.23\linewidth]{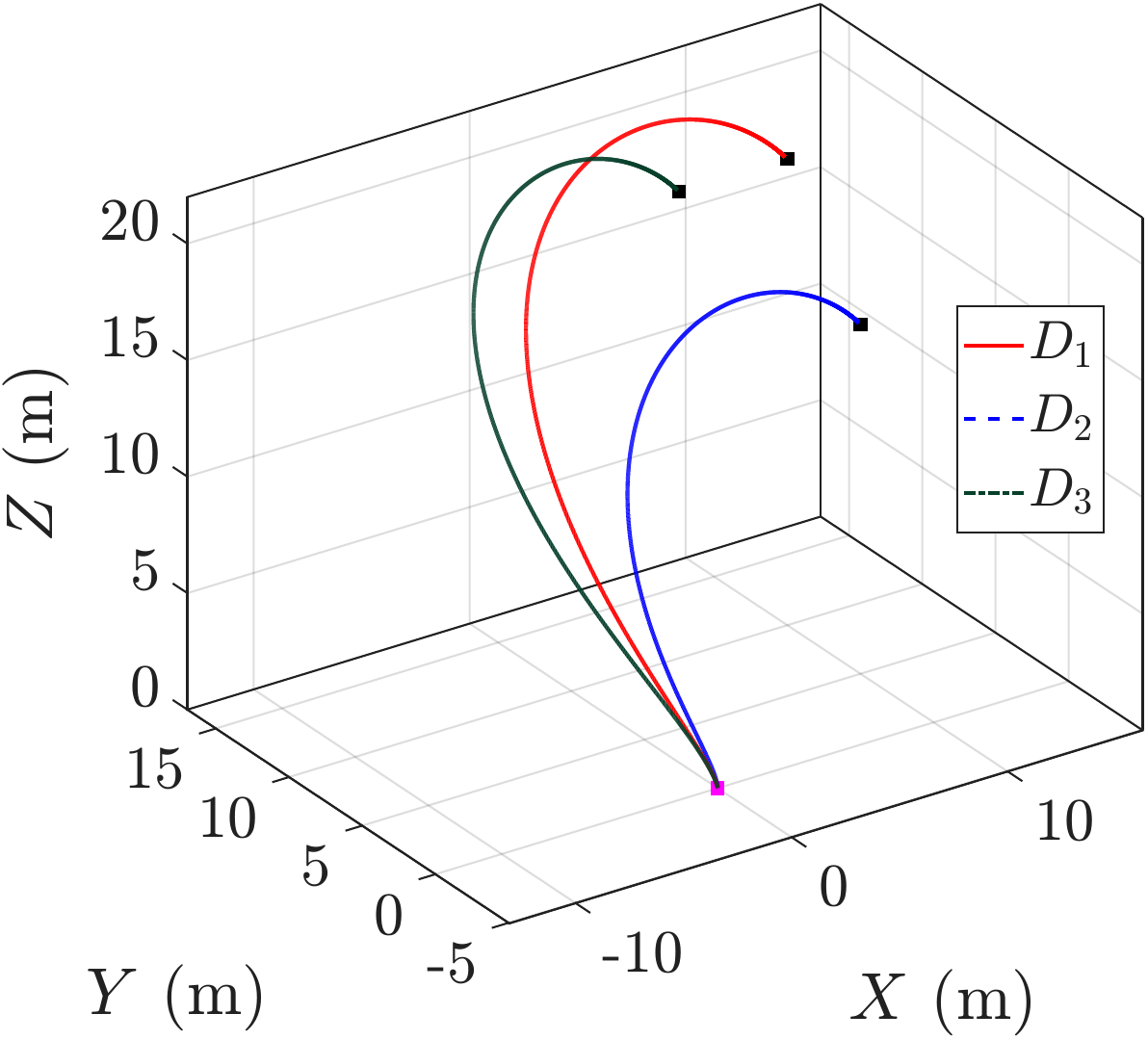}
			\label{fig:D_path}
		}%
		\subfloat[Speed, range and closing rate.]{
			\includegraphics[width=0.23\linewidth]{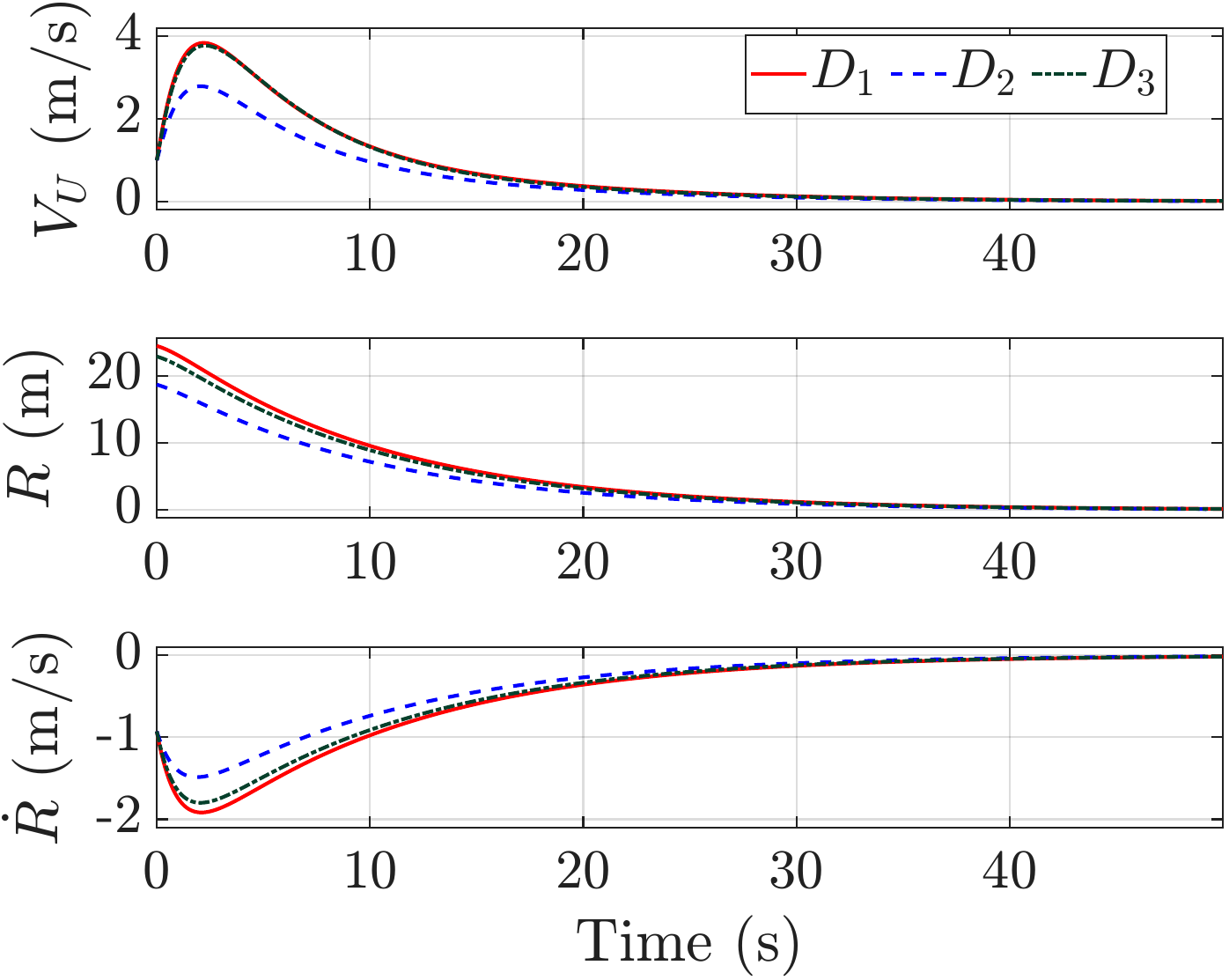}
			\label{fig:D_speed_range}
		}%
		\subfloat[Control acceleration.]{
			\includegraphics[width=0.23\linewidth]{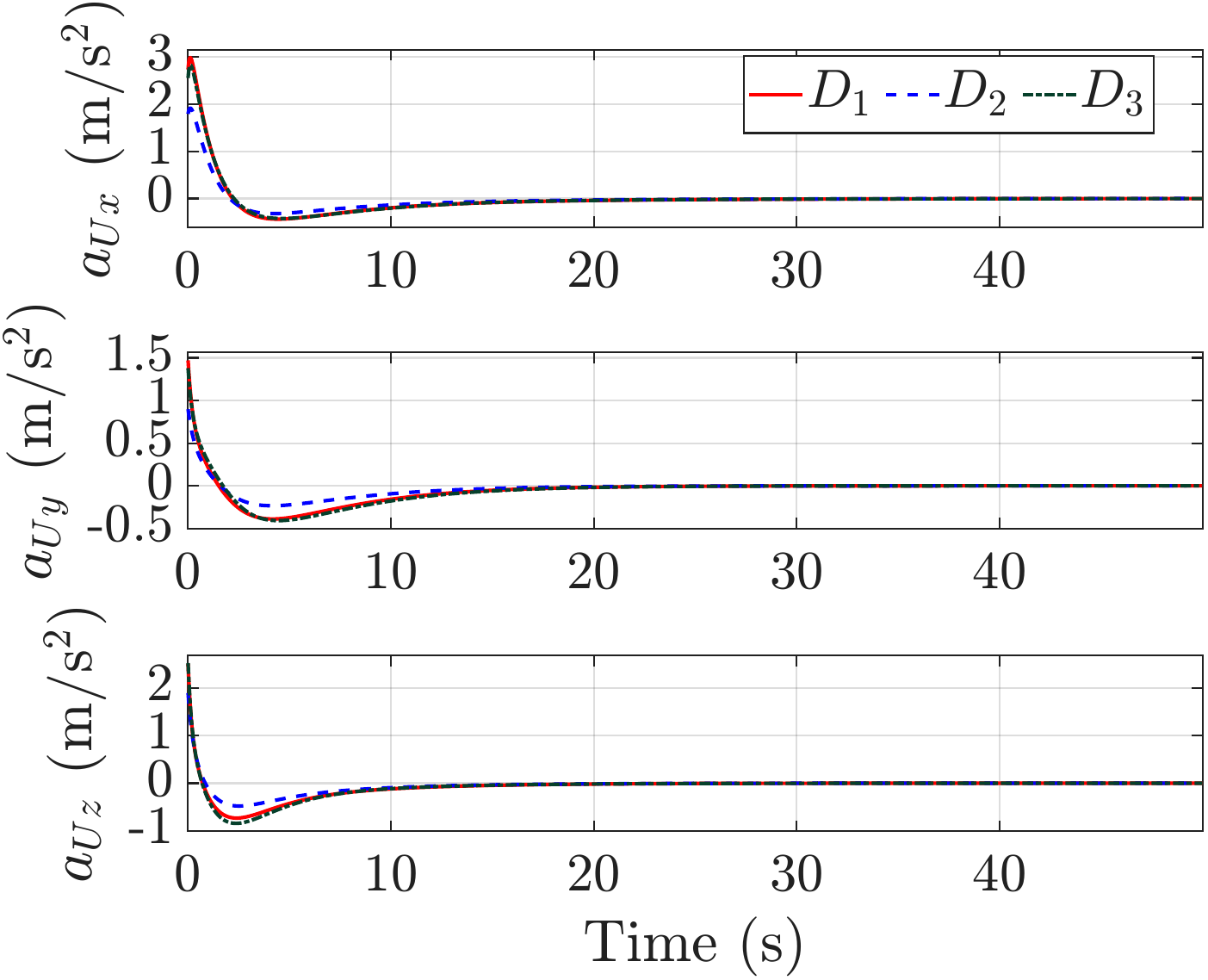}
			\label{fig:D_ctrl_accn}
		}%
		\subfloat[LOS angles and Lyapunov function.]{
			\includegraphics[width=0.23\linewidth]{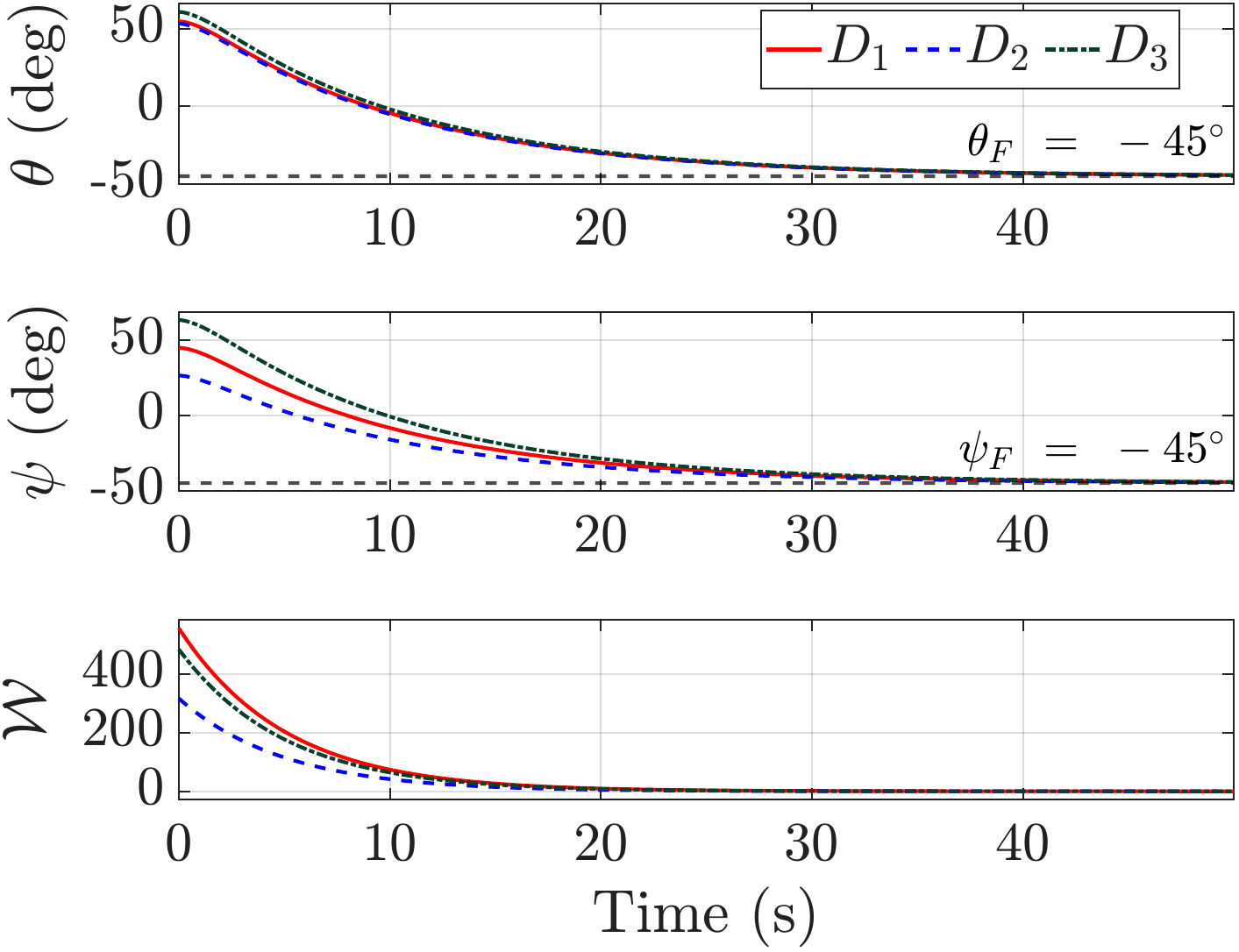}
			\label{fig:D_angles}
		}
		\caption{Demonstration of docking for various locations of the docking station while fixed vehicle initial location.}
		\label{fig:D}
	\end{figure*}

	Second, we consider a scenario when the DS is initialized with various locations denoted as $D_1:  [10,~ 10,~ 20,~-45^\circ,~-45^\circ]$, $D_2:  [10,~ 5,~ 15,~-45^\circ,~-45^\circ]$, and $D_3:  [5,~ 10,~ 20,~-45^\circ,~-45^\circ]$. The AV starts from the origin at a speed of $1$ m/s. The initial orientation of the velocity vector of the AV with respect to the LOS is $\psi_U = 10^\circ$ and $\theta_U = 20^\circ$. The performance for this scenario is illustrated in \cref{fig:D}.
	As depicted in \cref{fig:D_path}, the AV starts from the origin and successfully docks with the DS for all three cases $D_1$, $D_2$, and $D_3$, respectively. \cref{fig:D_speed_range} evidences that the AV applies control (see \cref{fig:D_ctrl_accn}) to reduce its speed and dock with the DS while maintaining the desired orientation (see \cref{fig:D_angles}).
	
	Finally, we validate the proposed docking strategy for docking with the DS at various approach angles. The AV is initialized with $[0,~ 0,~ 0,~ 1.0,~ 10^\circ,~ 20^\circ]$. The DS is located at ($10,~10,~10$) meters. The three cases of desired orientations of the docking are denoted with ($\theta_F,~\psi_F$) as $D_1:~ (10^\circ,~20^\circ)$, $D_2:~ (30^\circ,~60^\circ)$, and $ D_3:~ (60^\circ,~0^\circ)$.
	\begin{figure*}[!t]
		\centering
		\subfloat[Path of the vehicle.]{
			\includegraphics[width=0.23\linewidth]{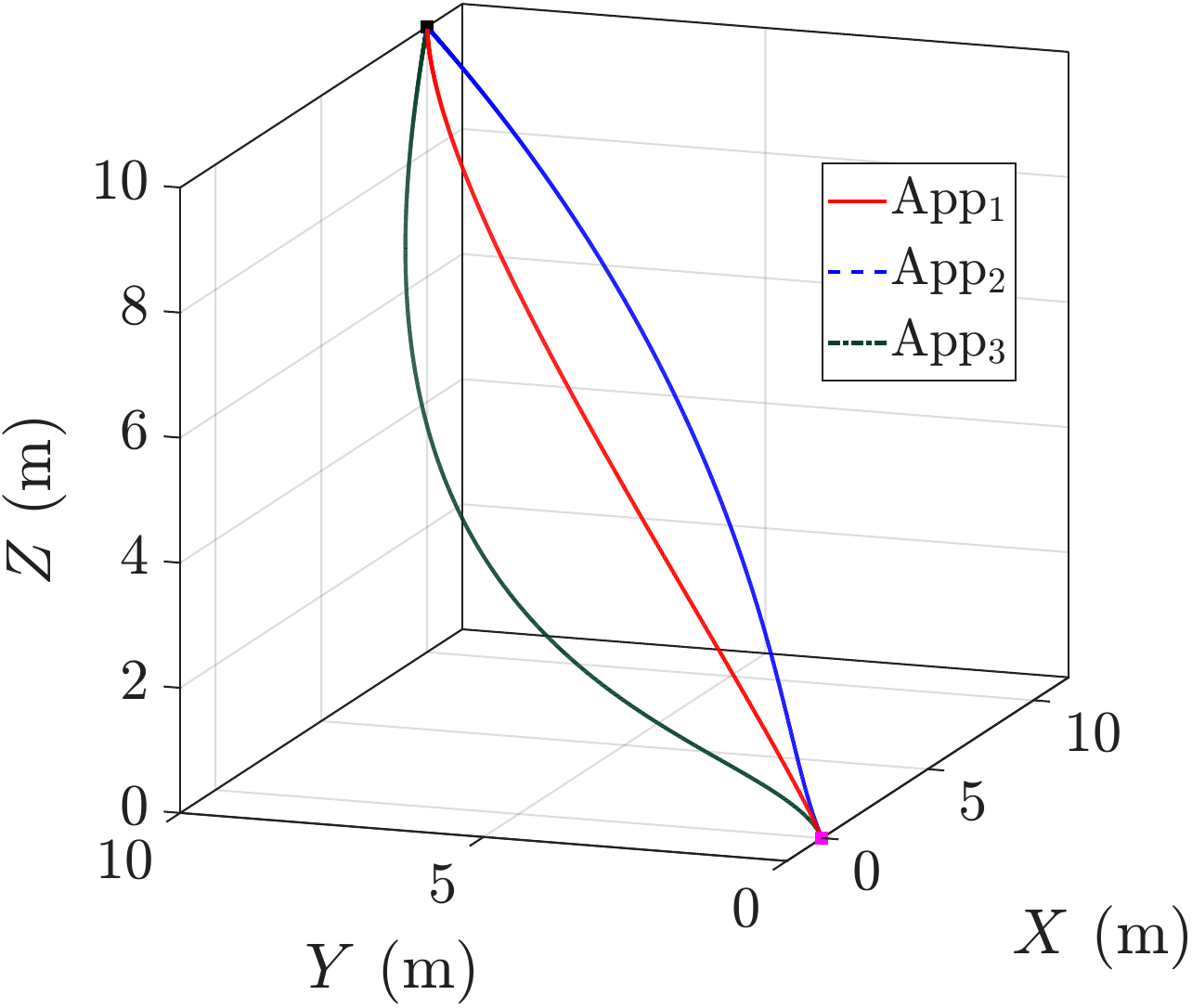}
			\label{fig:App_path}
		}%
		\subfloat[Speed, range and closing rate.]{
			\includegraphics[width=0.23\linewidth]{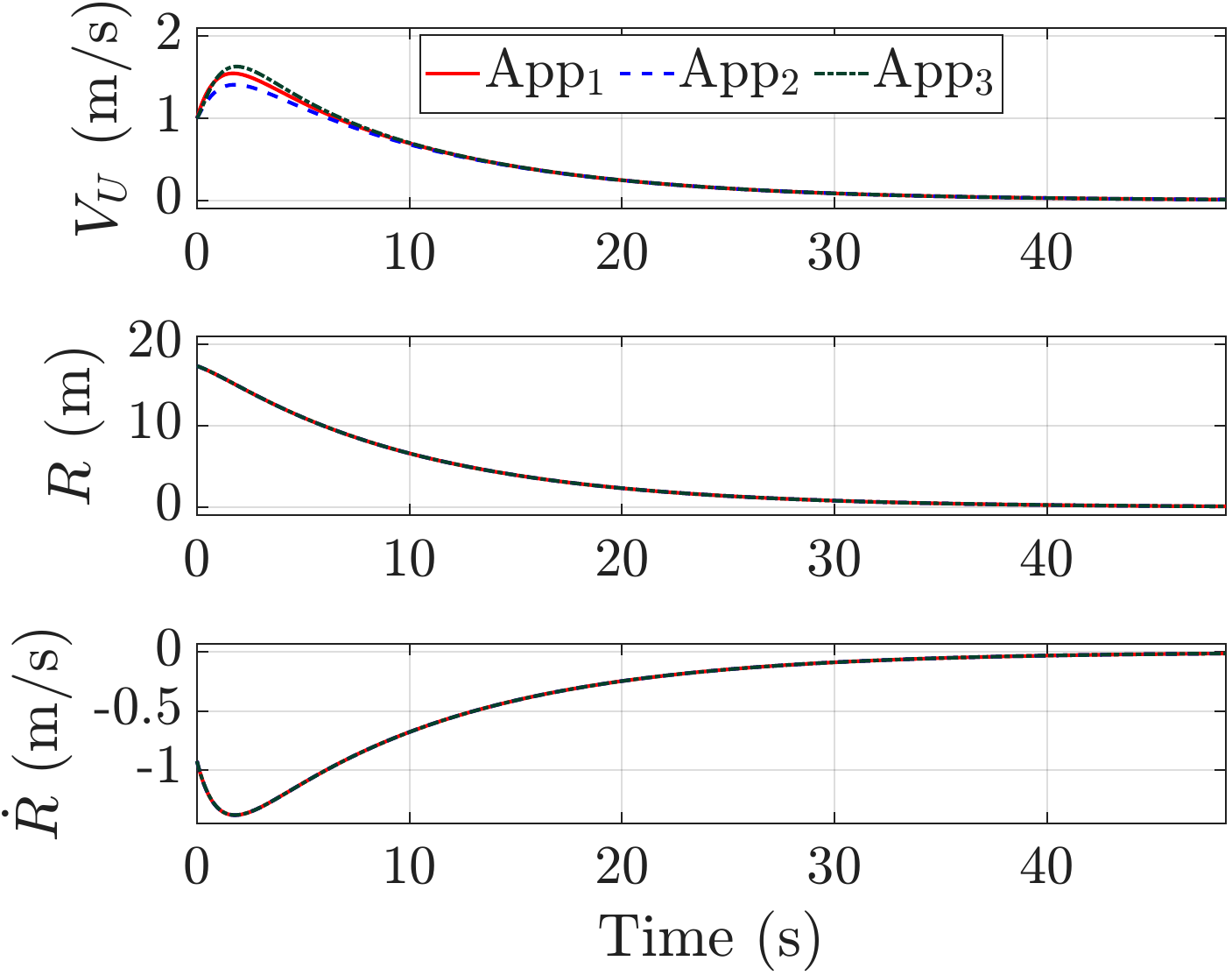}
			\label{fig:App_speed_range}
		}%
		\subfloat[Control acceleration.]{
			\includegraphics[width=0.23\linewidth]{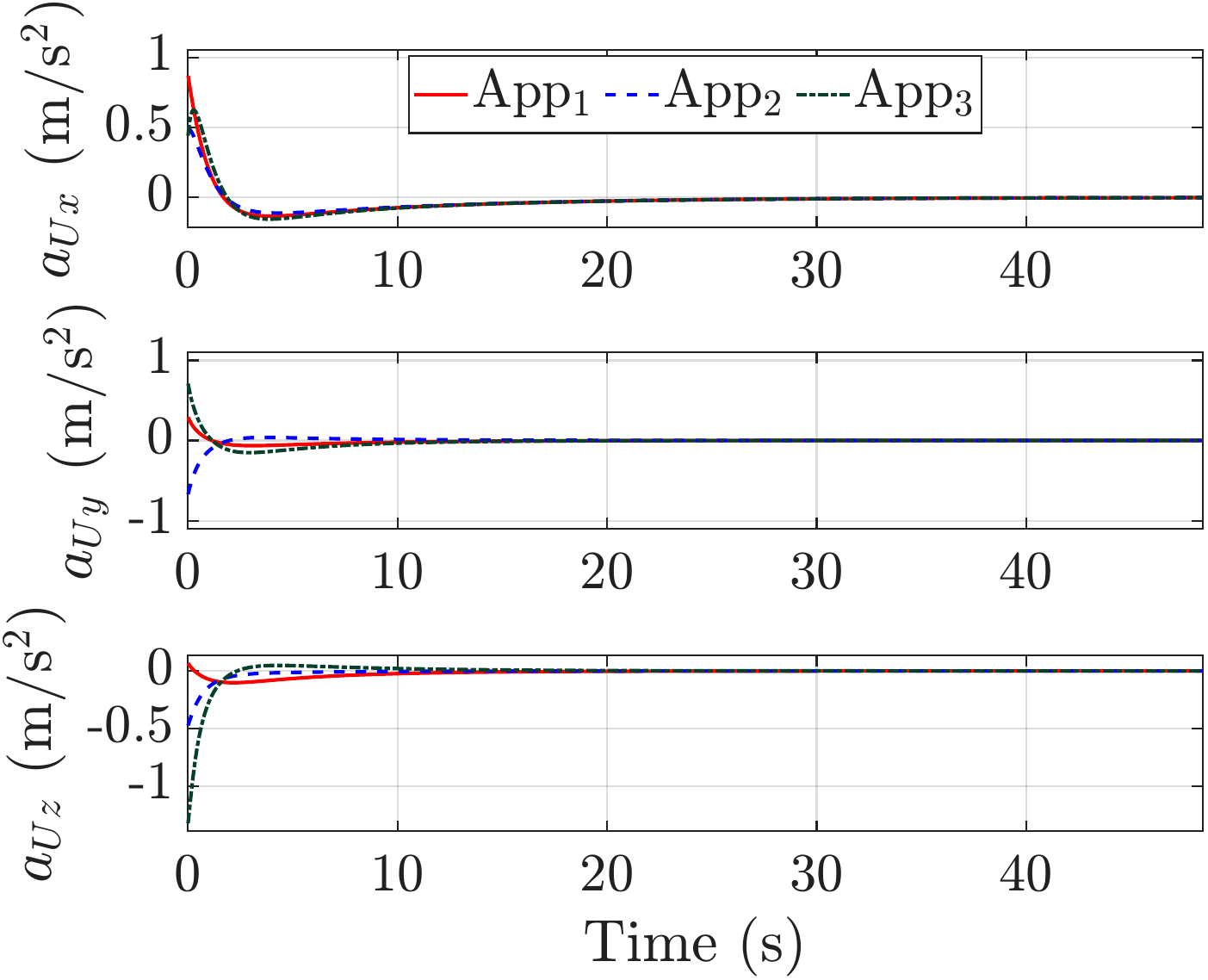}
			\label{fig:App_ctrl_accn}
		}%
		\subfloat[LOS angles and Lyapunov function.]{
			\includegraphics[width=0.23\linewidth]{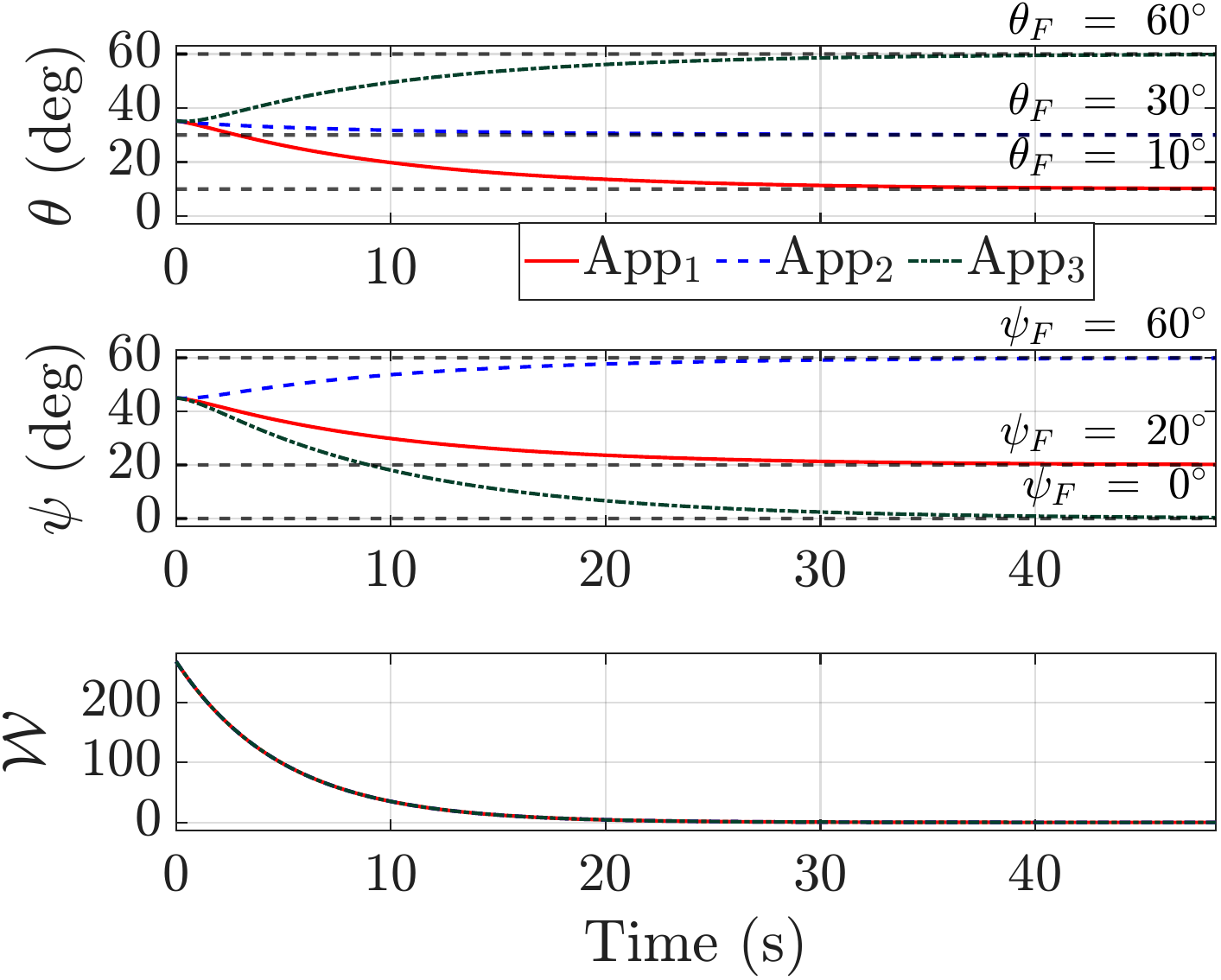}
			\label{fig:App_angles}
		}
		\caption{Demonstration of docking for various approach angles with the same docking station location and initial vehicle location.}
		\label{fig:App}
	\end{figure*}
	The AV starts from the origin and docks with the DS in all three orientations, denoted by $D_1$, $D_2$, and $D_3$, respectively, and the corresponding path of the AV is shown in \cref{fig:App_path}. Along with a speed reduction (refer \cref{fig:App_speed_range}), the AV also orients its velocity vector as desired (refer \cref{fig:App_angles}) to dock successfully. The control input profiles in \cref{fig:App_ctrl_accn} show a behavior similar to earlier cases.
	
	\section{Conclusion}
	In this work, we addressed the problem of autonomous docking with a stationary docking station located in a three-dimensional space. We formulated docking as a terminal-angle constraint problem and utilized the relative motion kinematics between the vehicle and the docking station to meet the spatial orientation and speed-reduction requirements for a successful docking. Unlike existing strategies, the proposed strategy presents a unified framework that does not require separate guidance laws for the homing and docking phases. This strategy relies only on range and LOS-related measurements for all phases of the mission, which reduces complexity and cost.
	Simulations demonstrated the working of the proposed strategy from various launch conditions and locations of the DS with different approach angles. This work can be potentially utilized for the docking of underwater vehicles, the landing of UAVs, the rendezvous and docking of spacecraft, and mobile robots. 
	
	\bibliography{references}
	
\end{document}